\pdfoutput=1
\documentclass[12pt,a4paper]{article}

\usepackage[T1]{fontenc}
\usepackage[utf8]{inputenc}
\usepackage{mathptmx}                 
\usepackage[margin=1in]{geometry}
\usepackage{amsmath,amssymb}
\usepackage{graphicx}
\usepackage{booktabs}
\usepackage{threeparttable}
\usepackage[font=small,labelfont=bf,labelsep=period,skip=6pt]{caption}
\usepackage{setspace}
\usepackage[round,authoryear]{natbib}
\usepackage{authblk}
\usepackage{xcolor}
\usepackage{microtype}
\usepackage{xurl}
\usepackage[colorlinks=true,linkcolor=black,citecolor=black,urlcolor=blue!60!black]{hyperref}

\IfFileExists{tfrupee.sty}{\usepackage{tfrupee}}{\providecommand{\rupee}{Rs\,}}

\newcommand{\KL}{D_{\mathrm{KL}}}
\newcommand{\DMF}{D_{\mathrm{KL}}(M\,\|\,F)}
\newcommand{\DFM}{D_{\mathrm{KL}}(F\,\|\,M)}
\newcommand{\pM}{p_{m}}
\newcommand{\pF}{p_{f}}
\newcommand{\dd}{\mathrm{d}}

\newif\ifanonymous
\anonymousfalse

\title{\textbf{An asymmetric gap: entropy, Kullback--Leibler divergence and the gender wage distribution in India's casual labor market}}

\ifanonymous
  \author{}
\else
  \author[1]{Rohita Gotety\thanks{Bachelor's student. Email: \href{mailto:rohita.gotety@gmail.com}{rohita.gotety@gmail.com}.}}
  \author[1]{Arpan Chakraborty\thanks{Assistant Professor, Corresponding author. Email: \href{mailto:arpan.ms97@gmail.com}{arpan.ms97@gmail.com}. ORCID: \href{https://orcid.org/0000-0002-7777-5643}{0000-0002-7777-5643}.}}
  \author[2]{Shourya Dutta\thanks{Email: \href{mailto:shouryadutta7777@gmail.com}{shouryadutta7777@gmail.com}. ORCID: \href{https://orcid.org/0000-0001-6867-2530}{0000-0001-6867-2530}.}}

  \affil[1]{Department of Economics, School of Humanities and Social Sciences, JAIN (Deemed-to-be University), Bengaluru, India}

  \affil[2]{Independent Researcher, Dhahran, Saudi Arabia}
\fi
\date{}

\begin{document}

\maketitle
\thispagestyle{empty}

\begin{abstract}
\noindent Official statistics suggest that the gender pay gap among India's casual workers has narrowed since the pandemic, as women's average daily wages rose faster than men's. Averages, however, can hide what happens across the wage range. Using Periodic Labour Force Survey (PLFS) data for 2022--2025, we compare the complete wage distributions of male and female casual workers in Delhi, Maharashtra, Tamil Nadu and West Bengal. Our tool is the Kullback--Leibler divergence, an information-theory measure of how poorly one distribution describes another, computed in both directions. To our knowledge, this is the first study based on PLFS data to connect entropy-based information theory, through the Kullback--Leibler divergence, with the asymmetry between male and female wage distributions in India's informal sector. The gap is large and persistent: at every rank, from the lowest-paid to the best-paid, women earn less than men. It is also lopsided. In every year, describing men's wages with women's loses 34--62 percent more information than the reverse, because women are concentrated in a narrow band of low wages that men also earn, while many men earn wages women rarely reach. This is consistent with a glass ceiling within casual work, though the data cannot reveal its causes. In 2023, as female participation surged, the direction most sensitive to low wages rose by about 20 percent, with and without survey weights. Results for 2025, the first year of a redesigned survey, depend on weighting. All estimates are descriptive; design-consistent uncertainty measures remain to be computed.


\end{abstract}

\newpage
\onehalfspacing

\section{Introduction}

Women's measured labor force participation in India rose markedly after the COVID-19 pandemic. According to the Periodic Labour Force Survey (PLFS), the share of women aged 15 and above who were working or looking for work (usual status) rose between 2022 and 2025 from 24.7 to 27.7 percent in urban areas and from 37.5 to 45.9 percent in rural areas \citep{mospi2026}. Over the same period, average daily earnings from casual labor other than public works rose from \rupee268 to \rupee315 for women and from \rupee424 to \rupee455 for men, so that women's average earnings rose from about 63 to about 69 percent of men's \citep{mospi2026}. On these figures, the gender pay gap in casual work narrowed.

Averages can mislead in two ways. First, the average gap can fall simply because a different group of women or men is working for wages, even if no individual woman is better paid than before \citep{mulligan2008,olivetti2008}. Second, an average says nothing about where the gap lies. Women may be held back at the bottom, because they are stuck in the lowest-paid work (a ``sticky floor''), or at the top, because they are kept out of the better-paid work (a ``glass ceiling''). The two problems call for different policies, yet they can produce the same average gap.

This paper looks at the whole range of wages with a tool from information theory, the Kullback--Leibler (KL) divergence \citep{shannon1948,kullback1951}. The idea can be put in everyday terms. Suppose that you know only how much women casual workers earn, and you are then shown the daily wage of a man chosen at random. If he earns \rupee400 a day, you will be surprised, because hardly any women earn that much. Now reverse the experiment: you know only men's wages and are shown a woman earning \rupee120 a day. You will be less surprised, because some men also earn wages in that range. The KL divergence measures this surprise, averaged over all workers, and so measures how poorly the wage pattern of one sex describes that of the other. The answer depends on the direction. We write $\DMF$ for the information lost when women's wages are used to describe men's, and $\DFM$ for the information lost in the reverse case.

Most applied work treats this difference between the two directions as a technical nuisance and averages it away. We argue that it is informative. $\DMF$ is large when many men earn wages that women rarely earn, which in casual work happens mainly at the top of the wage range, so it is most sensitive to a glass-ceiling pattern. $\DFM$ is large when many women earn wages that men rarely accept, which happens mainly at the bottom, so it is most sensitive to a sticky-floor pattern. The sum of the two directions tells us how big the gap is, and their difference tells us on which side of the wage range the two distributions overlap less. Neither direction is a pure measure of a ceiling or a floor, and neither explains why the distributions differ: together they show at which wage levels one group is common and the other is rare.

We apply this approach to casual wage laborers: workers who are paid by the day, or for short spells of work, on other people's farms and in other people's businesses, usually without job security or social security. Casual labor is the most precarious part of India's informal wage employment and accounted for 17.6 percent of female workers in 2025 \citep{mospi2026}. We study four large states spread across the north, west, south and east of the country, Delhi, Maharashtra, Tamil Nadu and West Bengal, which are also home to India's four traditional metropolitan cities. Using person-level PLFS data for 2022--2025 and the survey weights, we measure the gap in both directions in each year. A simple Gaussian benchmark illustrates how to read the result: if men's wages are more spread out than women's, so that women's wages are bunched into a narrower band, then the information lost in describing men's wages with women's is the larger of the two. Kolmogorov--Smirnov (KS) statistics and quantile--quantile (QQ) plots, two standard ways of comparing distributions, complement the analysis.

Four findings stand out. First, the gap is large and has not gone away. At every rank, from the lowest-paid worker to the best-paid, women earn less than men of the same rank, and at the wage level where the two groups differ most, the share of women earning that wage or less is 54 to 64 percentage points higher than the share of men. Second, the gap is lopsided in every year: describing men's wages with women's loses 34 to 62 percent more information than the reverse. Put differently, the wage of a typical man is 3.2 to 4.4 times as common among men as among women, whereas the wage of a typical woman is only 2.3 to 3.0 times as common among women as among men. Women are concentrated in a narrow band of low wages that men also earn, while many men earn wages that women rarely reach. This is consistent with a glass ceiling within casual work, although our data do not show what keeps women out of its better-paid segments. Third, 2023 stands out. In the year in which women's participation jumped, $\DFM$, the direction most sensitive to the bottom of the distribution, rose by about 20 percent, with or without the survey weights, while $\DMF$ did not rise. One possible explanation, which we cannot test with these data, is that many of the women who entered casual work took the lowest-paid jobs. Fourth, the results for 2025, the first year of a redesigned survey, depend on weighting: the weighted estimates show the gap shrinking by 16 percent, whereas the unweighted estimates show it growing. We therefore treat 2022--2024 as the main comparable period and 2025 as a separate check. Exploratory state-level estimates suggest that the gap differs considerably across the four states.

These findings are descriptive. The divergences do not yet come with design-consistent margins of error, they depend on how the wage distributions are smoothed, and they cannot separate discrimination from differences in skills, hours, industry or type of work. We therefore put most weight on patterns that hold in every year and with and without the survey weights, and we set out the remaining checks in Section~\ref{sec:conclusion}.

The paper makes three contributions. First, to the best of our knowledge, it is the first study based on PLFS data to connect entropy-based information theory, through the KL divergence, with the asymmetry of male and female wage distributions in India's informal sector. Second, it shows that the direction of the divergence, usually treated as a technical inconvenience, carries economic information. Quantile comparisons show how far apart men and women of the same rank are; the two directions of the divergence show how far the two groups occupy different wage levels at all, and whether this lack of overlap lies mainly at the top or at the bottom of the wage range. Third, it gives early evidence on how the gender gap in casual work has changed since the pandemic and shows that conclusions drawn from the redesigned PLFS depend on whether the survey weights are used. Section~\ref{sec:lit} reviews the literature, Section~\ref{sec:data} describes the data, Section~\ref{sec:framework} sets out the information-theoretic framework, Section~\ref{sec:results} presents the results, Section~\ref{sec:discussion} discusses them and Section~\ref{sec:conclusion} concludes.

\section{Literature review}\label{sec:lit}

\subsection{Glass ceilings and sticky floors}

Research on gender wage gaps across the wage distribution distinguishes two patterns. A glass ceiling is present when the gap widens at the top of the distribution, reflecting barriers to women's entry into the best-paid positions \citep{albrecht2003}. A sticky floor is present when the gap is widest at the bottom of the distribution. \citet{berheide1992} coined the term to describe the situation in the USA, where more than half of the women employed in state and local governments were ``stuck'' in low-paying, low-mobility and historically undervalued jobs in para-professional, administrative support and maintenance occupations. \citet{booth2003} defined a sticky floor as the situation in which otherwise identical men and women may be appointed to the same pay scale or rank, but women are appointed at the bottom of the scale while men are placed further up.

\citet{arulampalam2007} used quantile regression to examine gender pay gaps across the wage distribution in 11 European countries. They found gaps widening at the top in most countries, consistent with glass ceilings, and wide gaps at the bottom in some countries, consistent with sticky floors. \citet{ciminelli2021} found that, on average, sticky floors related to social norms, gender stereotyping and discrimination account for 40 percent of the gap among similarly qualified workers, a pattern that dominates in Central and Eastern European countries; the remaining 60 percent, a glass ceiling, was attributed to the motherhood penalty. \citet{gunewardena2008} examined whether the sticky floor in Sri Lanka depends on the sector of employment and on occupation. Using quantile regression decompositions, they found strong sticky floors and little evidence of glass ceilings, and because the sticky floors were not explained by occupational segregation or sectoral stratification (``sticky doors''), they reflect gaps within sectors and occupations. \citet{prakash2025} found sticky floors to be more common in lower-middle-income countries and glass ceilings in high-income and upper-middle-income countries. Evidence from other settings points in a similar direction. In Vietnam, \citet{nguyen2022} found that the gap is larger at higher quantiles and is mainly unexplained by worker characteristics. In Italy, \citet{bonanno2026} found that women experience higher wage inefficiency than men, which they interpret as structural frictions that prevent women from fully converting their skills into earnings, a pattern they link to sticky floor effects.

\subsection{Evidence from India}

Studies of regular wage and salaried workers in India largely find sticky floors. \citet{khanna2012} used the 2009--10 National Sample Survey and the Machado--Mata--Melly decomposition to identify a persistent sticky floor among regular workers. \citet{deshpande2018} confirmed this pattern for 1999--2000 and 2009--10, showing that the bulk of the gap was unexplained by observed characteristics and that the sticky floor became ``stickier'' over the decade. \citet{duraisamy2016} found a sticky floor in all segments of the labor market between 1983 and 2012: although the raw gender wage gap narrowed, the unexplained gap at the bottom of the distribution widened. \citet{agrawal2013} found sticky floors in urban areas but a glass ceiling in rural areas. The pattern is not confined to gender: \citet{deshpande2016} found larger caste gaps at the lower deciles of the earnings distribution of household enterprises, which they interpret as a sticky floor.

Casual labor has received less attention. \citet{mahajan2017} showed that higher female labor supply accounts for a large part of the wider gender gap in agricultural wages in southern India, highlighting the number of women seeking work as an important determinant of casual wage gaps. Most directly relevant to this paper, \citet{trivedi2025} applied quantile regression and the Machado--Mata--Melly decomposition to PLFS data for 2020--21. For regular salaried workers, the gap fell from 0.709 log points (103.2 percent) at the 10th percentile to 0.160 log points (17.4 percent) at the 90th, a clear sticky floor. For casual workers, by contrast, the raw gap rose from 41.8 percent at the 10th percentile to 64.5 percent at the 90th, and its large unexplained component suggests that women face a different form of disadvantage in this segment.

\subsection{Composition, selection and the measured gap}

The changing population of women in paid work is a central challenge in interpreting changes in the gender gap. \citet{mulligan2008} showed that much of the narrowing of the gender gap in the USA reflected a change in the composition of the female workforce towards women with higher earning potential. \citet{olivetti2008} showed that countries with low female employment tend to have small measured gaps because working women are positively selected, so that observed gaps understate underlying inequality; \citet{blau2017} review this evidence. Selection can also run the other way, when women with lower earning potential enter paid work. In India, \citet{abraham2022} found that women were more likely than men to lose work during the 2020 lockdown and less likely to return; men often moved into self-employment and daily wage work, whereas women had few such options and many left the labor force. \citet{klasen2015} observed that the participation of married women in urban India stagnated between 1987 and 2011 despite economic growth and rising education, which suggests that recent entry into the workforce is likely to be concentrated in specific segments of work. Across 125 countries, \citet{saha2025} find that labor market conditions shape how far women's education translates into participation, and that employer discrimination can limit the participation even of educated women.

Decomposition methods address part of this problem. The Oaxaca--Blinder decomposition divides mean gaps into explained and unexplained components \citep{oaxaca1973,blinder1973}, and quantile-based methods extend it across the distribution \citep{machado2005,melly2005,firpo2009}. These methods require a model of the wage-generating process and a well-measured set of covariates. A complementary approach, used in this paper, compares the full unconditional distributions directly, with tools that make no assumptions about their shape. Such comparisons cannot attribute the gap to particular causes, but they can show whether the two distributions moved closer together and which of their features changed. Information theory provides such tools, as the next subsection discusses.

\subsection{Information theory and the measurement of inequality}

Information theory entered economics through the measurement of inequality. \citet{theil1967} defined his inequality index as the divergence between the distribution of income shares and that of population shares, a special case of the relative entropy introduced by \citet{shannon1948} and \citet{kullback1951}. The generalized entropy family of \citet{cowell1980} and \citet{shorrocks1980} extends this idea into the standard class of additively decomposable inequality measures, and \citet{maasoumi1993} surveys the wider use of entropy and divergence measures in economics and econometrics. \citet{kullback1959} developed the divergence as a measure of the ``discrimination information'' available for distinguishing between two hypotheses, and \citet{jeffreys1946} proposed its symmetric version. Symmetric measures such as the Jeffreys divergence and the Jensen--Shannon divergence \citep{lin1991} are often preferred because they treat the two distributions alike. In the study of gender gaps, however, the two distributions are not interchangeable: men's and women's wages are shaped by different access to work, and which group's distribution fails to describe the other's is itself informative. The divergence complements, rather than replaces, quantile-based comparisons and decompositions: quantile methods compare men and women at the same rank, whereas the divergence measures how far the two groups occupy different wage levels, and in which direction. We are not aware of studies that use the directional asymmetry of the KL divergence to characterize gender wage gaps, and in particular none based on the PLFS.

\section{Data}\label{sec:data}

\subsection{Data and sample}

The data come from the Periodic Labour Force Survey (PLFS), conducted by the National Statistics Office (NSO) of the Ministry of Statistics and Programme Implementation (MoSPI), Government of India. The PLFS has been India's principal source of labor force statistics since 2017. It measures labor force participation and employment in the usual status (principal and subsidiary activity) and in the current weekly status, in rural and urban areas. We use the person-level unit records for the calendar years 2022, 2023, 2024 and 2025. The 2025 data come from a redesigned survey, introduced in January 2025, which expanded the sample from about 102,000 to about 272,000 households and introduced a monthly rotational panel in both rural and urban areas; MoSPI cautions that estimates for 2025 may not be strictly comparable with those for earlier years \citep{mospi2026}. We therefore treat 2022--2024 as the main comparable period and report the 2025 results separately.

The analysis covers four states: Delhi, Maharashtra, Tamil Nadu and West Bengal. For brevity, we refer to the National Capital Territory of Delhi as a state. The four are spread across the north, west, south and east of the country; they range from an almost entirely urban territory (Delhi) to states with large rural labor markets; and they contain India's four traditionally recognized metropolitan cities (Delhi, Mumbai, Chennai and Kolkata), which were also the four largest urban agglomerations in the 2011 Census. The samples are state-wide and include both rural and urban areas, so the results describe casual labor in these four states as a whole, not in the metropolitan cities alone, and they are not representative of India. Extending the analysis to all states is a natural next step.%
\footnote{The pooled sample for 2022--2025 includes 22,210 men and 8,592 women: 4,573 men and 1,709 women in 2022, 4,386 men and 1,708 women in 2023, 4,289 men and 1,633 women in 2024, and 8,962 men and 3,542 women in 2025.}

Workers are classified by their current weekly status (CWS), which refers to the seven days preceding the date of the survey. A person is counted as working in the CWS if he or she worked for at least one hour on at least one day during that week. A worker's CWS status is the principal activity in the reference week, determined by a priority-cum-major-time criterion: working takes priority over seeking work, and among work activities the status is the one in which the person spent the most time. The sample consists of persons whose principal activity status in the CWS was casual wage labor, defined in the PLFS as casual engagement in others' farm or non-farm enterprises, household or non-household, in return for a wage according to the terms of a daily or periodic work contract. We use the CWS because the PLFS records daily wage earnings for the reference week of the CWS, so that the activity status and the wage refer to the same period. This choice excludes persons who did some casual work during the reference week but whose principal weekly activity was something else, such as self-employment or domestic duties. Because women are more likely than men to combine short spells of casual work with other activities, the sample may under-represent women's occasional casual work. Casual wage labor also includes work in public works, including work under the Mahatma Gandhi National Rural Employment Guarantee Act (MGNREGA), which guarantees at least 100 days of wage employment in a financial year to every rural household whose adult members volunteer to do unskilled manual work. Wages in public works are set administratively and may follow different rules from private casual work. Because the PLFS identifies public works and MGNREGA work by separate status codes, results that exclude them are a natural robustness check, which we leave for a future version; our sample therefore differs from the MoSPI earnings series cited in the introduction, which refers to casual labor other than public works. The wage variable is, for each worker, the median of the daily wage earnings from casual labor recorded in the time-disposition schedule for the reference week, in rupees. Because all comparisons are made between men and women within a year, nominal wages need not be deflated: both summary measures used below are invariant to a common rescaling of the two distributions. Exploratory analysis used box plots and estimated probability density functions of male and female wages. All computations were carried out in R.

Estimates are reported in two forms. The weighted estimates apply the survey sampling weights provided with the PLFS unit-level data, so that each respondent counts in proportion to the number of persons he or she represents in the four states. Because the PLFS is a stratified multi-stage sample with unequal selection probabilities, the weighted estimates are the appropriate basis for statements about the population of casual workers, and we treat them as our main results. The unweighted estimates treat every sampled person equally and describe the wage distributions of the respondents in the sample; we report them for comparison. Both sets of estimates are computed with the same procedures and differ only in whether the survey weights are applied.

\section{An information-theoretic framework}\label{sec:framework}

\subsection{Entropy, cross-entropy and relative entropy}

Let $\pM$ and $\pF$ denote the densities of daily wages of male and female casual workers, and let $p$ and $q$ denote any two such densities. The (differential) Shannon entropy of $p$ and the cross-entropy of $p$ relative to $q$ are \citep{shannon1948,cover2006}
\begin{equation}\label{eq:entropy}
H(p)=-\int_{0}^{\infty}p(x)\log p(x)\,\dd x,\qquad
H(p,q)=-\int_{0}^{\infty}p(x)\log q(x)\,\dd x .
\end{equation}
The KL divergence, or relative entropy, of $q$ from $p$ is the difference between the two \citep{kullback1951,joyce2011}:
\begin{equation}\label{eq:kl}
\KL(p\,\|\,q)=\int_{0}^{\infty}p(x)\log\frac{p(x)}{q(x)}\,\dd x=H(p,q)-H(p).
\end{equation}
Equation~\eqref{eq:kl} has two interpretations that we use throughout. In coding terms, if wages drawn from $p$ are described with a code that is optimal for $q$, $\KL(p\,\|\,q)$ is the expected excess description length per observation, in nats when logarithms are natural \citep{cover2006}. In statistical terms, it is the expected log-likelihood ratio,
\begin{equation}\label{eq:llr}
\KL(p\,\|\,q)=\mathbb{E}_{p}\bigl[\log\Lambda(X)\bigr],\qquad \Lambda(x)=\frac{p(x)}{q(x)},
\end{equation}
so that $\exp\{\KL(p\,\|\,q)\}$ is the geometric-mean likelihood ratio: the factor by which a typical observation from $p$ is more likely under $p$ than under $q$. By the Chernoff--Stein lemma, $\KL(p\,\|\,q)$ is also the best achievable exponential rate at which the type II error of a test of $p$ against $q$ falls with the sample size $n$, when the type I error is held fixed \citep{cover2006}:
\begin{equation}\label{eq:stein}
\lim_{n\to\infty}-\frac{1}{n}\log\beta_{n}=\KL(p\,\|\,q).
\end{equation}

The divergence is non-negative, equals zero only when $p=q$ (Gibbs' inequality) and is finite only if $q$ is positive wherever $p$ is. It is invariant under any one-to-one transformation applied to both distributions \citep{kullback1951}, so it takes the same value for wages and for log wages, and deflating both by a common price index leaves it unchanged. It is jointly convex in $(p,q)$ \citep{cover2006}. And it is not symmetric: in general $\KL(p\,\|\,q)\neq\KL(q\,\|\,p)$.

\subsection{Two directions: where the distributions fail to overlap}

Applied to the two sexes, the divergence has two directions,
\begin{equation}\label{eq:directions}
\DMF=\int_{0}^{\infty}\pM(x)\log\frac{\pM(x)}{\pF(x)}\,\dd x,\qquad
\DFM=\int_{0}^{\infty}\pF(x)\log\frac{\pF(x)}{\pM(x)}\,\dd x .
\end{equation}
$\DMF$ is the information lost when women's wage distribution is used to describe men's wages; $\DFM$ is the information lost in the reverse case. Suppose, as Figures~\ref{fig:dens_pooled} and~\ref{fig:dens_year} show for the data, that the two densities cross once, at a wage $x^{*}$, with $\pF>\pM$ below $x^{*}$ and $\pM>\pF$ above it. Then each direction splits into a positive and a negative part:
\begin{align}
\DMF&=\underbrace{\int_{x^{*}}^{\infty}\pM(x)\log\frac{\pM(x)}{\pF(x)}\,\dd x}_{\text{ceiling term, }C_{M}\ge 0}\;-\;\int_{0}^{x^{*}}\pM(x)\log\frac{\pF(x)}{\pM(x)}\,\dd x, \label{eq:ceiling}\\
\DFM&=\underbrace{\int_{0}^{x^{*}}\pF(x)\log\frac{\pF(x)}{\pM(x)}\,\dd x}_{\text{floor term, }C_{F}\ge 0}\;-\;\int_{x^{*}}^{\infty}\pF(x)\log\frac{\pM(x)}{\pF(x)}\,\dd x. \label{eq:floor}
\end{align}
All positive contributions to $\DMF$ come from wages above $x^{*}$, where men are present and women are scarce: the term $C_{M}$ grows as the male upper tail extends into wage ranges that women rarely reach. All positive contributions to $\DFM$ come from wages below $x^{*}$, where women are present and men are scarce: the term $C_{F}$ grows as women crowd into low wages that men rarely accept. Neither direction is a pure ceiling or floor measure, however. Each combines a positive term from one side of the crossing with a negative term from the other, and local crossings created by wage heaping make the split only approximate. What equations~\eqref{eq:ceiling}--\eqref{eq:floor} show is where each direction gets its weight: $\DMF$ is most sensitive to the upper part of the wage range and $\DFM$ to the lower part. We therefore refer to them as the ceiling-sensitive and the floor-sensitive directions. We use ``ceiling'' and ``floor'' descriptively, for the regions in which the two distributions fail to overlap; the divergences do not by themselves identify the barriers, or the discrimination, that the terms glass ceiling and sticky floor usually refer to. We do not estimate $C_{M}$ and $C_{F}$ separately in this paper; doing so, and checking how the split depends on $x^{*}$, is a priority for future work.

\subsection{Size and structure of the gap}

We summarize the two directions by their sum and their difference. The Jeffreys divergence \citep{jeffreys1946},
\begin{equation}\label{eq:jeffreys}
J=\DMF+\DFM=\int_{0}^{\infty}\bigl(\pM(x)-\pF(x)\bigr)\log\frac{\pM(x)}{\pF(x)}\,\dd x ,
\end{equation}
is symmetric and measures the size of the gap. The asymmetry,
\begin{equation}\label{eq:asym}
\Delta=\DMF-\DFM,\qquad A=\frac{\Delta}{J}\in(-1,1),
\end{equation}
measures its structure: $\Delta>0$ means that the male distribution is harder to describe with the female one than the reverse, which in the single-crossing case indicates that the mismatch in the upper part of the wage range outweighs the mismatch in the lower part. Using equation~\eqref{eq:kl}, the asymmetry can be written entirely in terms of entropies,
\begin{equation}\label{eq:asym_entropy}
\Delta=\bigl[H(\pM,\pF)-H(\pF,\pM)\bigr]-\bigl[H(\pM)-H(\pF)\bigr],
\end{equation}
which separates the difference in cross-entropies from the difference in the intrinsic uncertainty (entropy) of the two wage distributions.

\subsection{A Gaussian benchmark}\label{sec:benchmark}

To illustrate what drives the sign of $\Delta$, consider a stylized benchmark in which log wages are normally distributed, $\log W_{g}\sim\mathcal{N}(\mu_{g},\sigma_{g}^{2})$ for $g\in\{m,f\}$. Because the divergence is invariant to the log transformation, it can be computed for log wages with the standard formula for the divergence between two normal distributions \citep{cover2006}. Writing $r=\sigma_{m}/\sigma_{f}$ for the dispersion ratio and $z=(\mu_{m}-\mu_{f})/\sigma_{f}$ for the mean gap in units of the female standard deviation, the two directions are
\begin{equation}\label{eq:gauss_dir}
\DMF=-\ln r+\frac{r^{2}+z^{2}}{2}-\frac{1}{2}, \qquad \DFM=\ln r+\frac{1+z^{2}}{2r^{2}}-\frac{1}{2},
\end{equation}
so that
\begin{align}
J&=\frac{r^{2}+r^{-2}}{2}-1+\frac{z^{2}}{2}\bigl(1+r^{-2}\bigr), \label{eq:gauss_J}\\
\Delta&=\underbrace{\frac{r^{2}-r^{-2}}{2}-2\ln r}_{g(r)}+\frac{z^{2}}{2}\bigl(1-r^{-2}\bigr). \label{eq:gauss_delta}
\end{align}
Since $g(1)=0$ and $g'(r)=(r^{2}-1)^{2}/r^{3}\ge0$, the first term in equation~\eqref{eq:gauss_delta} has the sign of $r-1$, and so does the second term whenever $z\neq0$. Hence $\Delta>0$ if and only if $r>1$: $\DMF$ exceeds $\DFM$ exactly when male log wages are more dispersed than female log wages. The mean gap $z$ raises both directions and magnifies the asymmetry, but only the dispersion ratio determines its sign. This is the formal counterpart of equations~\eqref{eq:ceiling}--\eqref{eq:floor}: when women's wages are compressed into a narrower band than men's, the male distribution spills into regions where the female density is thin.

Equations~\eqref{eq:gauss_J}--\eqref{eq:gauss_delta} can also be read in reverse. Given estimates of $J$ and $\Delta$, they are two equations in the two unknowns $r$ and $z$, which we solve numerically to translate each pair of estimated divergences into a ``Gaussian-equivalent'' dispersion ratio and mean gap. These are illustrative diagnostics, not estimates of the wage-generating process: observed casual wages are skewed, heaped at round values and not log-normal, heaping adds to both directions of the divergence, and other two-parameter families would map the same divergences into different parameters. We use them only to translate the divergences into more familiar quantities. As a check, we compare the maximum gap between the two cumulative distributions implied by the Gaussian-equivalent parameters with the observed KS statistic (Appendix~\ref{app:calib}).

\subsection{Estimation}

Each density is estimated with a Gaussian kernel density estimator, using R's default bandwidth \citep{silverman1986}. In the weighted estimates the kernel estimator uses the survey weights $w_{i}$,
\begin{equation}\label{eq:kde}
\hat{p}_{g}(x)=\frac{\sum_{i\in g}w_{i}\,K_{h}(x-X_{i})}{\sum_{i\in g}w_{i}},\qquad K_{h}(u)=\frac{1}{h\sqrt{2\pi}}\exp\Bigl(-\frac{u^{2}}{2h^{2}}\Bigr),
\end{equation}
and the unweighted estimates set $w_{i}=1$. Both densities are evaluated on a common grid $\{x_{k}\}$ and normalized to probabilities $p_{k}$ and $q_{k}$, and each direction is computed as the discrete analogue
\begin{equation}\label{eq:kld}
\hat{D}_{\mathrm{KL}}(p\,\|\,q)=\sum_{k}p_{k}\log\frac{p_{k}+\varepsilon}{q_{k}+\varepsilon},\qquad \varepsilon=10^{-10},
\end{equation}
with natural logarithms, so that divergences are in nats (1 nat $\approx$ 1.443 bits). The constant $\varepsilon$ avoids taking the logarithm of zero where an estimated density underflows; individual terms can be negative, but the divergence as a whole is non-negative. The two directions are computed from the same estimated densities, with the roles of $p$ and $q$ exchanged. Plug-in estimates of the divergence are biased in finite samples: because the divergence is jointly convex, sampling noise in the estimated densities biases them upwards (by Jensen's inequality), whereas kernel smoothing tends to bias them downwards, and both effects depend on the sample size, the bandwidth and the discretization of the support \citep{wang2005}. We therefore focus on comparisons across years and between the two directions, which are computed in the same way, rather than on absolute levels. Because both wage distributions are heaped at round values, the kernel estimates smooth over the heaps, and the divergences are sensitive to wage ranges in which one estimated density is close to zero. This version uses a single specification. We have not yet examined the sensitivity of the results to the bandwidth (including under- and oversmoothing), the kernel, the evaluation grid and the constant $\varepsilon$, or compared them with a discrete estimator on the observed wage support, and the results should be read with this in mind.

\subsection{Complementary distributional tests}

We complement the divergences with the two-sample KS test \citep{kolmogorov1933,smirnov1939,dodge2008}. With empirical distribution functions $F_{1}(x)=n_{1}^{-1}\sum_{i}\mathbf{1}\{X_{1i}\le x\}$ for men and $F_{2}(x)=n_{2}^{-1}\sum_{j}\mathbf{1}\{X_{2j}\le x\}$ for women (weighted in the weighted estimates), the statistic is
\begin{equation}\label{eq:ks}
D=\sup_{x}\,\bigl|F_{1}(x)-F_{2}(x)\bigr| ,
\end{equation}
the largest difference, across all wage levels, between the shares of men and women earning at or below a given wage. Under simple random sampling, the null hypothesis of identical distributions is rejected at level $\alpha$ when $D>c(\alpha)\sqrt{(n_{1}+n_{2})/(n_{1}n_{2})}$, with $c(0.05)\approx1.358$. The PLFS, however, is a stratified multistage sample with unequal selection probabilities, and wages are heaped, so these conventional critical values and p-values are not design-valid. We use $D$ as a descriptive statistic and report the conventional p-values only for reference (Appendix Table~\ref{tab:ks}). $D$ is most sensitive to the body of the distribution and comparatively insensitive to the tails \citep{mason1983}, which makes it a useful contrast to the tail-sensitive divergences. QQ plots \citep{wilk1968}, which plot the female quantile $Q_{f}(\tau)$ against the male quantile $Q_{m}(\tau)$ for $\tau\in(0,1)$, show where the gap lies (Appendix~\ref{app:figs}).

\section{Results}\label{sec:results}

All the estimates below are point estimates. Design-consistent standard errors for the divergences have not yet been computed, so statements that one estimate is larger than another, or that a measure rose or fell, are descriptive rather than tests of significance. We put most weight on patterns that hold in every year and with and without the survey weights, which are less likely to reflect sampling noise or a particular specification, and we treat 2022--2024 as the main comparable period.

\subsection{The shape of the two distributions}

\begin{figure}[htbp]
\centering
\includegraphics[width=0.85\textwidth]{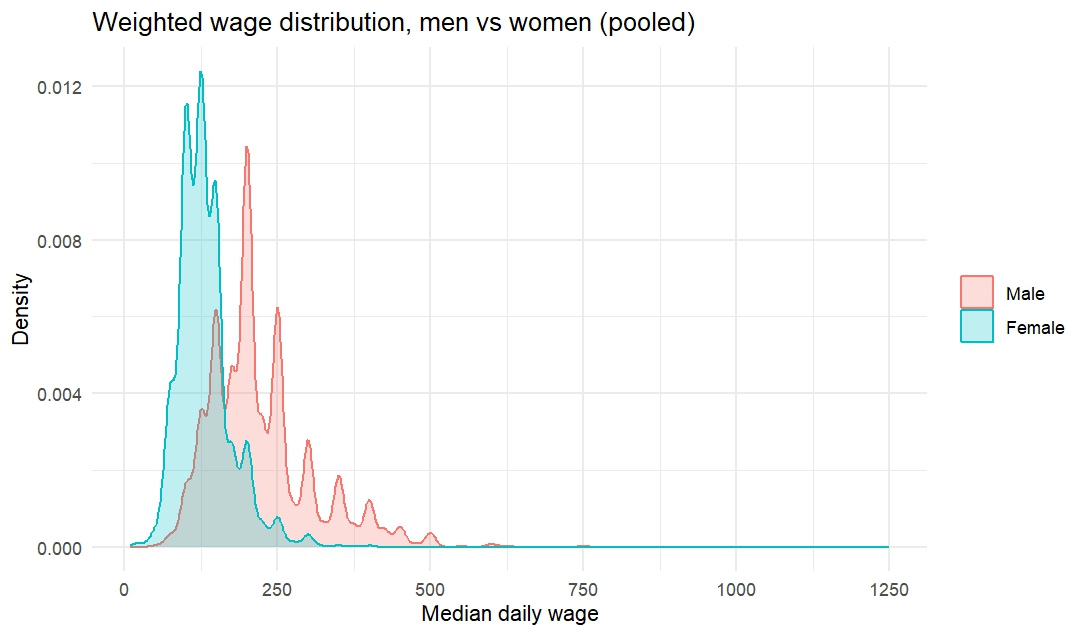}
\caption{Survey-weighted densities of median daily casual wages, men and women, 2022--2025 pooled}\label{fig:dens_pooled}
\begin{minipage}{0.85\textwidth}\footnotesize
\textit{Notes:} Gaussian kernel density estimates with survey weights; wages in rupees per day.\\
\textit{Source:} Authors' calculations based on PLFS person-level data.
\end{minipage}
\end{figure}

\begin{figure}[htbp]
\centering
\includegraphics[width=0.9\textwidth]{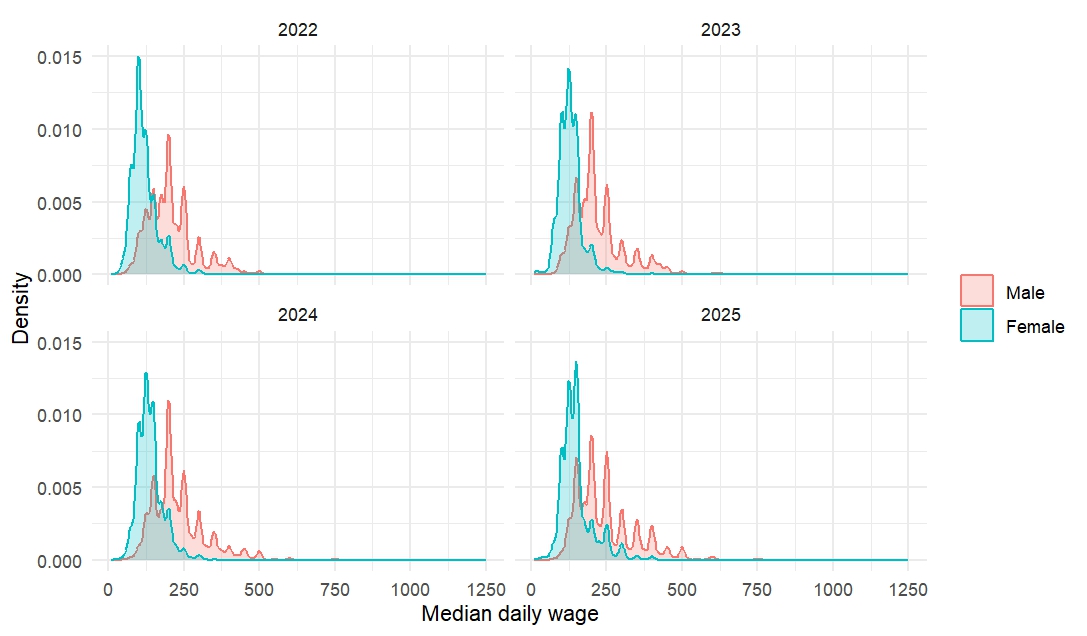}
\caption{Survey-weighted densities of median daily casual wages, men and women, by year}\label{fig:dens_year}
\begin{minipage}{0.9\textwidth}\footnotesize
\textit{Notes:} As for Figure~\ref{fig:dens_pooled}.\\
\textit{Source:} Authors' calculations based on PLFS person-level data.
\end{minipage}
\end{figure}

Figure~\ref{fig:dens_pooled} shows the survey-weighted wage densities for the pooled sample, and Figure~\ref{fig:dens_year} shows them for each year. Two features frame everything that follows. First, women's wages are concentrated in a narrow, low band, with most of the female mass between about \rupee75 and \rupee175, whereas men's wages peak at about \rupee200 and \rupee250 and have a long upper tail, with heaps at round values up to about \rupee500. Apart from small local crossings created by heaping, the densities cross once, at roughly \rupee150--175, which is the single-crossing configuration of equations~\eqref{eq:ceiling}--\eqref{eq:floor}. Second, both distributions are heaped at round wage levels, a familiar feature of daily wage reporting. Over time, the main female peak moves up, from about \rupee100 in 2022 to about \rupee150 in 2025. The unweighted densities show the same features (Appendix~\ref{app:udens}).

\subsection{A large and asymmetric information gap}

\begin{figure}[htbp]
\centering
\includegraphics[width=\textwidth]{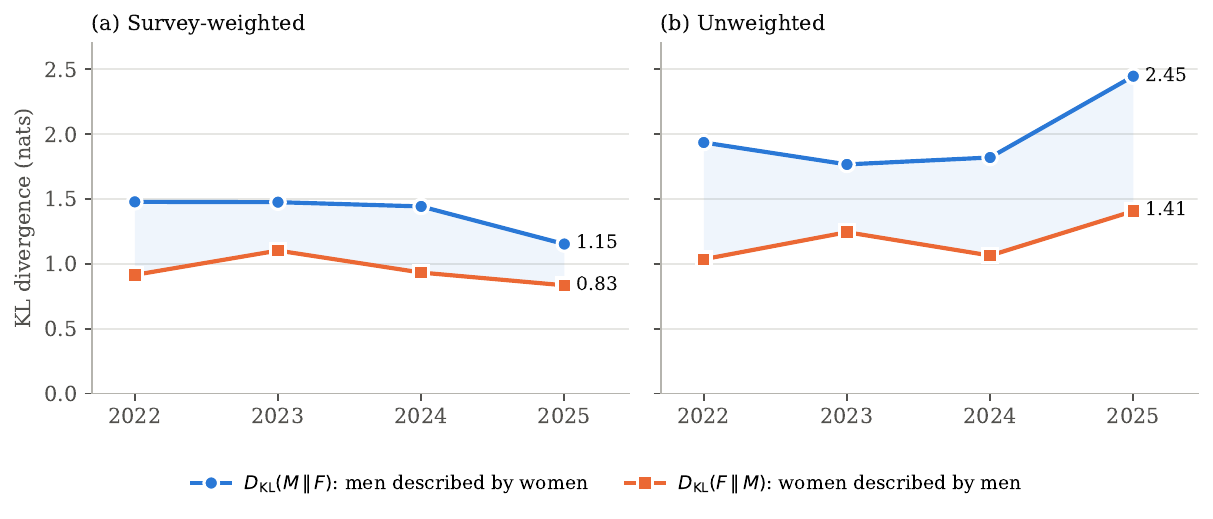}
\caption{Directional Kullback--Leibler divergences between male and female daily casual wage distributions, 2022--2025}\label{fig:kl}
\begin{minipage}{\textwidth}\footnotesize
\textit{Notes:} Divergences in nats. The shaded band is the asymmetry $\Delta=\DMF-\DFM$.\\
\textit{Source:} Authors' calculations based on PLFS person-level data.
\end{minipage}
\end{figure}

Table~\ref{tab:main} reports the survey-weighted divergences in both directions, together with the Jeffreys divergence, the asymmetry and the KS statistic, and Figure~\ref{fig:kl} plots the two directions over time. Three results stand out.

\begin{table}[htbp]
\centering
\begin{threeparttable}
\caption{Directional Kullback--Leibler divergences between the male and female distributions of daily casual wages, survey-weighted, 2022--2025}\label{tab:main}
\small
\begin{tabular}{@{}lcccccc@{}}
\toprule
 & \multicolumn{2}{c}{Directional divergence} & Size & \multicolumn{2}{c}{Structure} & KS \\
\cmidrule(lr){2-3}\cmidrule(lr){4-4}\cmidrule(lr){5-6}\cmidrule(l){7-7}
Year & $\DMF$ & $\DFM$ & $J$ & $\Delta$ & $\DMF/\DFM$ & $D$ \\
\midrule
2022 & 1.478 & 0.915 & 2.393 & 0.563 & 1.62 & 0.565 \\
2023 & 1.475 & 1.102 & 2.578 & 0.373 & 1.34 & 0.638 \\
2024 & 1.443 & 0.933 & 2.376 & 0.510 & 1.55 & 0.577 \\
2025 & 1.153 & 0.835 & 1.987 & 0.318 & 1.38 & 0.535 \\
\addlinespace
2022--2025 (pooled) & 1.293 & 0.964 & 2.257 & 0.329 & 1.34 & 0.580 \\
\bottomrule
\end{tabular}
\begin{tablenotes}[flushleft]
\footnotesize
\item \textit{Notes:} Divergences in nats. $\DMF$ is the information lost when the female wage distribution is used to describe male wages; $\DFM$ is the reverse. $J=\DMF+\DFM$ (Jeffreys divergence) and $\Delta=\DMF-\DFM$. $D$ is the weighted two-sample Kolmogorov--Smirnov statistic; the conventional p-values, which are not adjusted for the survey design, are below 0.001 in every year (Appendix Table~\ref{tab:ks}). The 2025 data come from the redesigned PLFS and are not strictly comparable with earlier years. Pooled values are not directly comparable with annual values. Unweighted estimates are in Appendix Table~\ref{tab:unweighted}.
\item \textit{Source:} Authors' calculations based on PLFS person-level data.
\end{tablenotes}
\end{threeparttable}
\end{table}

First, the gap is large. The Jeffreys divergence lies between 1.99 and 2.58 nats, and the KS statistic implies that, at the point of greatest separation, the share of women earning at or below a given wage exceeds that of men by 54 to 64 percentage points. In likelihood terms (equation~\ref{eq:llr}), a typical male wage is $e^{1.478}\approx4.4$ times as likely under the male distribution as under the female one in 2022, and about 3.2 times in 2025.

Second, the gap is asymmetric in every year. $\DMF$ exceeds $\DFM$ by 34 to 62 percent, and $\Delta$ is positive throughout. A typical female wage is only 2.3 to 3.0 times as likely under the female distribution as under the male one, compared with 3.2 to 4.4 times for a typical male wage under the male distribution. In other words, women's wages are reasonably well described by the male distribution, because men also work at low casual wages, but men's wages are poorly described by the female distribution, because women are largely absent from the upper part of the casual wage range. In terms of equations~\eqref{eq:ceiling}--\eqref{eq:floor}, the mismatch on the ceiling side outweighs the mismatch on the floor side. The density plots show the compression of the female distribution directly (Figures~\ref{fig:dens_pooled} and~\ref{fig:dens_year}), and the illustrative Gaussian benchmark of Section~\ref{sec:benchmark} expresses it in familiar terms: the Gaussian-equivalent dispersion ratio $r$ lies between 1.16 and 1.28, which corresponds to male log wages being about 16 to 28 percent more dispersed than female log wages, and the Gaussian-equivalent mean gap $z$ lies between 1.50 and 1.70 female standard deviations (Appendix Table~\ref{tab:calib}). These parameters reproduce the observed KS statistics fairly closely (0.51--0.57 implied against 0.54--0.64 observed), even though they are derived only from the two divergences. A location--dispersion reading therefore captures much of the difference, although it cannot capture the heaping and multiple peaks visible in Figures~\ref{fig:dens_pooled} and~\ref{fig:dens_year}.

Third, the asymmetry does not depend on the survey weights. It holds in every year in the unweighted data as well, where $\DMF$ exceeds $\DFM$ by 42 to 87 percent (Appendix Table~\ref{tab:unweighted}). It also holds in fifteen of the sixteen weighted state-year estimates, the exception being Maharashtra in 2024 (Appendix Table~\ref{tab:states_w}), although the state-level estimates are less reliable (Section~\ref{sec:states}).

Divergences measured in nats are unfamiliar to most economists, so three reference points may help in reading their size. First, if male and female log wages were normal with equal variances and means $d$ standard deviations apart, the Jeffreys divergence would equal $d^{2}$. On this rough yardstick, the weighted values of $J$ correspond to distributions whose centers are 1.4 to 1.6 standard deviations apart; two such distributions would share only about 42 to 48 percent of their probability mass and would have a KS statistic of 0.52 to 0.58, close to the observed 0.54 to 0.64. Second, the QQ plots (Appendix~\ref{app:figs}) show that female quantiles are roughly two-thirds or more of the corresponding male quantiles in the lower part of the distribution and about one-half to three-fifths of them where male quantiles lie between \rupee300 and \rupee500. Third, the official all-India ratio of female to male average casual earnings, which excludes public works, rose from about 0.63 to 0.69 over the period \citep{mospi2026}. A Jeffreys divergence of about 2.4 nats thus describes a gap in which women at a given rank earn roughly one-half to two-thirds of what men at the same rank earn, and in which men and women largely work at different wage levels.

\subsection{2023: a rise in the floor-sensitive direction}

The two directions moved differently in 2023. $\DFM$ rose by 20.5 percent in the weighted data (from 0.915 to 1.102) and by 20.1 percent in the unweighted data (from 1.036 to 1.244), while $\DMF$ was unchanged in the weighted data and fell by 8.7 percent in the unweighted data. The whole increase in the Jeffreys divergence in 2023 (from 2.393 to 2.578) therefore came from the female-to-male direction, and the asymmetry fell from 0.563 to 0.373. In terms of equation~\eqref{eq:floor}, this is what one would expect if more of the female distribution moved to low wages that men rarely earn. The Gaussian-equivalent parameters are consistent with this, with the mean gap essentially unchanged ($z$ moved from 1.68 to 1.70) and the dispersion ratio falling from 1.28 to 1.16 as female log wages became more dispersed relative to male log wages. The KS statistic, which is most sensitive to the body of the distribution, reached its highest value in 2023 under both weighting schemes (0.638 weighted, 0.641 unweighted). The rise in $\DFM$ is also widespread: in the weighted state-level estimates it rose between 2022 and 2023 in all four states, by 22 to 26 percent in West Bengal, Maharashtra and Tamil Nadu and from a very low level in Delhi (Appendix Table~\ref{tab:states_w}).

This is the one change in the period that appears with and without the survey weights, although without standard errors we cannot say whether it is statistically significant. Its timing is suggestive. Female participation rose sharply between 2022 and 2023, by 2.5 percentage points in urban areas and 9.8 percentage points in rural areas \citep{mospi2026}. One possible explanation is that new entrants took up the lowest-paid casual work: this would raise the floor-sensitive direction, as observed, while leaving the ceiling-sensitive direction, which depends mainly on the upper part of the male distribution, largely unaffected. \citet{mahajan2017} find that higher female labor supply is associated with wider gender gaps in agricultural wages, which is consistent with this channel. We cannot test the explanation here, because the PLFS cross-sections used in this paper do not identify new entrants or track their wages. Comparing the composition of female casual workers in 2022 and 2023 by age, education, rural or urban location, industry, occupation and state would be a first step. In 2024, $\DFM$ fell back to 0.933, close to its 2022 level.

\subsection{2025: a separate check under the redesigned survey}

In the weighted data, both directions fell in 2025: $\DMF$ by 20.1 percent and $\DFM$ by 10.5 percent, so that the Jeffreys divergence fell by 16.3 percent to 1.987, its lowest value in the period, and the KS statistic fell to 0.535. The Gaussian-equivalent mean gap narrowed from 1.67 to 1.50 female standard deviations, and the main female peak in Figure~\ref{fig:dens_year} moved up to about \rupee150. Taken alone, these would be signs of partial convergence, driven mainly by the ceiling-sensitive direction. The unweighted data, however, show the opposite: both directions rose by about a third (by 34 and 32 percent), and the unweighted Jeffreys divergence is almost twice the weighted one in 2025 (3.85 against 1.99), whereas it exceeded the weighted one by 17 to 24 percent in earlier years. The PLFS was redesigned in January 2025, with a much larger sample, a new rotational panel and new weights \citep{mospi2026}, so the gap between the weighted and unweighted estimates may reflect changes in the composition of the sample, the rotational design or the weights. We cannot distinguish among these explanations here. We therefore do not interpret the weighted decline as evidence of convergence; it needs to be confirmed with later rounds of the redesigned survey.

\subsection{Heterogeneity across states}\label{sec:states}

The state-level estimates (Appendix Tables~\ref{tab:states_w} and~\ref{tab:states_u}) are exploratory. They rest on much smaller samples and are likely to be more sensitive to the density estimation than the aggregate estimates. This is clearest for Delhi, where the female sample is very small: its unweighted estimate for 2023 is undefined, and its other unweighted estimates reverse the asymmetry with implausibly large values of $\DFM$. With these caveats, the states appear to differ. Tamil Nadu has by far the largest divergences, with a weighted Jeffreys divergence between 4.36 and 4.83 in every year and no sign of convergence; $\DMF$ exceeds $\DFM$ in every year, by 26 to 78 percent. West Bengal is the opposite case: its weighted divergences are much lower in 2024 and 2025 than in 2022 and 2023, its Jeffreys divergence falling from 2.41 in 2022 to 0.95 in 2025, and its unweighted divergences also fell. Delhi has the most asymmetric weighted estimates, with $\DMF$ between 2.4 and 5.0 times $\DFM$, but these estimates are volatile. Maharashtra shows no clear trend and the only reversal of the asymmetry in the weighted data, in 2024. A minimum sample-size rule and design-consistent standard errors are needed before these differences can be interpreted.

\section{Discussion}\label{sec:discussion}

\subsection{What the entropy lens adds}

Average wage ratios compress the gender gap into a single number. Quantile comparisons, such as QQ plots and quantile decompositions, compare men and women at the same rank of their own distributions and show how far apart they are at each rank. The directional divergences answer a different question: how far the two groups occupy different wage levels at all. They weigh each wage level by how common it is in one group and how rare it is in the other, so they are largest when one group is concentrated where the other is almost absent. This has three practical advantages. First, the two directions show which group's wage levels the other fails to cover, and they can move independently, as in 2023, when the floor-sensitive direction rose while the ceiling-sensitive direction did not. Second, the divergence has a statistical interpretation that quantile gaps lack: by equation~\eqref{eq:llr} and the Chernoff--Stein lemma, it measures how easily one could tell from wages alone whether a sample of workers came from the male or the female distribution. Third, because the divergence is invariant to monotone transformations of wages, it does not depend on whether wages are measured in rupees or in logs, or on the price level, whereas the size of a quantile gap does. Inequality indices such as the Theil index, by contrast, measure dispersion within a distribution, and their between-group components compare only group means. The divergences complement these tools rather than replace them: they show where the two distributions differ, not why.

\subsection{A ceiling-type pattern in casual work}

Studies of regular salaried workers in India consistently find sticky floors \citep{khanna2012,deshpande2018,duraisamy2016}. Among casual laborers in the four states, the evidence points the other way: the ceiling-sensitive direction dominates, the female distribution is compressed relative to the male one, and the QQ plots show a proportional gap that is larger in the upper half of the distribution than at the bottom (Appendix~\ref{app:figs}). This matches the national finding of \citet{trivedi2025} that the gender gap among casual workers rises towards the top of the distribution. We read this as evidence of a gendered distributional disadvantage that is consistent with a glass-ceiling pattern, not as evidence of a particular mechanism. Several explanations are possible, and our unconditional comparisons cannot separate them: occupational segregation within casual work, differences in hours worked per day, differences in industry and regional composition, differences in education and skills, unequal access to particular types of casual work, and discrimination in pay for similar work. The 2023 episode shows that a floor-side mismatch can be added to this pattern when many women enter casual work at once.

\subsection{Composition}

The directional view also sharpens the question of composition. A change in the mix of women in casual work, such as entry at the bottom, should move the floor-sensitive direction first, whereas a change in women's access to better-paid tasks should move the ceiling-sensitive direction. The 2023 rise in $\DFM$ is consistent with the former; the weighted 2025 fall in $\DMF$, if it is confirmed, would be consistent with the latter. Testing these interpretations directly requires reweighting the female sample to hold its composition by rural or urban sector, state, industry, occupation, age and education constant over time.

\section{Conclusion}\label{sec:conclusion}

This paper asked whether the gender pay gap among casual workers in four large Indian states narrowed after the pandemic, as rising average wages suggest. Instead of comparing averages, we compared the whole wage distributions of men and women with the KL divergence, a measure of how poorly one distribution describes another, and we measured it in both directions. The gap remains large, and it is lopsided. Women's wages are concentrated in a narrow band of low wages in which men also work, while a large share of men earn wages that women rarely reach, so describing men's wages with women's loses much more information than the reverse. This pattern is consistent with a glass ceiling within casual work, although the data do not show its causes. The only change that appears with and without the survey weights is a 20 percent rise in the floor-sensitive direction in 2023, the year in which women's participation jumped. The results for 2025, the first year of the redesigned survey, depend on weighting and are not interpreted as convergence. All of these findings are descriptive and await design-consistent measures of uncertainty.

\subsection{Practical and social implications}

Three practical lessons follow, with the caveat that our results show where the gap lies, not why. First, average wage ratios are not enough to judge whether women's pay is catching up with men's. The two directional measures used here can be computed from the published survey data and show whether a change in the gap is happening mainly at the bottom or at the top of the wage range. Second, the Code on Wages, 2019, which came into force with India's other labor codes on 21 November 2025, prohibits discrimination on the ground of gender in wages for the same work or work of a similar nature and extends minimum wage protection to all workers \citep{goi2019}. Our results suggest that monitoring under the Code should look both at the low daily rates that many women are paid and at whether women obtain the better-paid casual jobs at all. Third, if women's near absence from better-paid casual work reflects barriers to entering particular tasks, measures that widen access to such work, such as training and skill certification, could help to narrow the gap. Our data cannot show whether skills, rather than segregation, hours or discrimination, are the binding constraint, and this needs to be tested before such policies are prioritized. Because casual workers have little job security or social protection, these gaps matter directly for household welfare and for women's economic security.

\subsection{Limitations and future research}

The analysis has several limitations, which also set out the next steps. First, the estimates are descriptive. The divergences carry no standard errors, and the conventional KS p-values ignore the stratification, clustering and unequal selection probabilities of the PLFS; a bootstrap that respects the survey design is needed for the two directions, the Jeffreys divergence, the asymmetry and their changes over time. Second, the divergences come from a single density specification. Because wages are heaped at round values, their sensitivity to the bandwidth, the kernel, the evaluation grid and the constant $\varepsilon$ needs to be examined, together with a comparison with a discrete estimator on the observed wage support. Third, the ceiling and floor terms of equations~\eqref{eq:ceiling}--\eqref{eq:floor} are not estimated separately, and the Gaussian-equivalent parameters are illustrative; a decomposition of the Jeffreys divergence into location, dispersion and shape components that does not rely on normality would strengthen the interpretation. Fourth, the comparisons are unconditional and cannot separate discrimination from differences in education, experience, hours or type of work; distributional decomposition methods \citep{machado2005,firpo2009}, corrections for selection into casual work \citep{heckman1979,olivetti2008} and reweighting to hold the composition of the female sample constant would address this. Fifth, the sample definition matters: it uses principal weekly status, includes public works and MGNREGA, and covers four states, so results that exclude public works, results that include persons who did casual work as a secondary activity, and an extension to all states are natural robustness checks. Finally, the 2025 results depend on weighting and may be affected by the survey redesign, and the state-level estimates rest on small samples, particularly for Delhi.

\section*{Declarations}

\noindent\textbf{Conflict of interest.} The authors declare no conflict of interest.

\smallskip
\noindent\textbf{Funding.} This research received no specific grant from any funding agency in the public, commercial or not-for-profit sectors.

\smallskip
\noindent\textbf{Data availability.} The PLFS unit-level data are publicly available from the microdata portal of the Ministry of Statistics and Programme Implementation (\url{https://microdata.gov.in}). The R code used for the analysis is available from the corresponding author on request.

\smallskip
\noindent\textbf{Use of generative AI.} Claude Opus 5.5 (Anthropic) was used to edit parts of the text. Claude Sonnet 5.5 (Anthropic) was used to generate part of the R code for preparing the data. Perplexity AI was used to compile a preliminary list of sources for the literature review. The authors reviewed all text, code, and results and take full responsibility for the content.

\begingroup
\singlespacing
\small

\endgroup

\clearpage
\appendix
\renewcommand{\thetable}{\thesection\arabic{table}}
\renewcommand{\thefigure}{\thesection\arabic{figure}}
\renewcommand{\theequation}{\thesection\arabic{equation}}
\renewcommand{\theHtable}{app.\thetable}
\renewcommand{\theHfigure}{app.\thefigure}
\renewcommand{\theHequation}{app.\theequation}
\setcounter{table}{0}
\setcounter{figure}{0}

\section{Gaussian-equivalent parameters}\label{app:calib}

Table~\ref{tab:calib} reports the values of $(r,z)$ obtained by solving equations~\eqref{eq:gauss_J}--\eqref{eq:gauss_delta} for each pair of estimated divergences, together with the maximum gap between the two cumulative distributions implied by these parameters, $\sup_{x}|\Phi(x)-\Phi((x-z)/r)|$, and the observed KS statistic. The implied and observed values are close in the weighted data and move together across years, with the highest value in 2023 and the lowest in 2025. The parameters are illustrative diagnostics (Section~\ref{sec:benchmark}), not estimates of the wage-generating process.

\begin{table}[htbp]
\centering
\begin{threeparttable}
\caption{Gaussian-equivalent dispersion ratio and mean gap implied by the directional divergences}\label{tab:calib}
\small
\begin{tabular}{@{}lccccc@{}}
\toprule
 & Dispersion ratio & Mean gap & \multicolumn{2}{c}{Max.\ CDF gap} & \\
\cmidrule(lr){4-5}
Year & $r=\sigma_{m}/\sigma_{f}$ & $z$ & Implied & Observed $D$ & \\
\midrule
\multicolumn{6}{@{}l}{\textit{Survey-weighted}}\\
2022 & 1.276 & 1.678 & 0.545 & 0.565 & \\
2023 & 1.158 & 1.704 & 0.572 & 0.638 & \\
2024 & 1.247 & 1.665 & 0.546 & 0.577 & \\
2025 & 1.177 & 1.499 & 0.512 & 0.535 & \\
2022--2025 (pooled) & 1.159 & 1.593 & 0.542 & 0.580 & \\
\addlinespace
\multicolumn{6}{@{}l}{\textit{Unweighted}}\\
2022 & 1.378 & 1.901 & 0.585 & 0.572 & \\
2023 & 1.193 & 1.861 & 0.606 & 0.641 & \\
2024 & 1.314 & 1.860 & 0.585 & 0.591 & \\
2025 & 1.324 & 2.168 & 0.654 & 0.592 & \\
2022--2025 (pooled) & 1.192 & 1.751 & 0.578 & 0.595 & \\
\bottomrule
\end{tabular}
\begin{tablenotes}[flushleft]
\footnotesize
\item \textit{Notes:} $r$ and $z$ solve equations~\eqref{eq:gauss_J}--\eqref{eq:gauss_delta} for the estimated $J$ and $\Delta$; $z$ is the male--female gap in mean log wages in units of the female standard deviation. The implied gap is $\sup_{x}|\Phi(x)-\Phi((x-z)/r)|$, where $\Phi$ is the standard normal distribution function.
\item \textit{Source:} Authors' calculations.
\end{tablenotes}
\end{threeparttable}
\end{table}

\clearpage
\section{Additional tables}\label{app:tables}
\setcounter{table}{0}

\begin{table}[htbp]
\centering
\begin{threeparttable}
\caption{Directional Kullback--Leibler divergences, unweighted, 2022--2025}\label{tab:unweighted}
\small
\begin{tabular}{@{}lcccccc@{}}
\toprule
Year & $\DMF$ & $\DFM$ & $J$ & $\Delta$ & $\DMF/\DFM$ & KS $D$ \\
\midrule
2022 & 1.936 & 1.036 & 2.971 & 0.900 & 1.87 & 0.572 \\
2023 & 1.767 & 1.244 & 3.011 & 0.522 & 1.42 & 0.641 \\
2024 & 1.820 & 1.064 & 2.884 & 0.756 & 1.71 & 0.591 \\
2025 & 2.447 & 1.407 & 3.853 & 1.040 & 1.74 & 0.592 \\
\addlinespace
2022--2025 (pooled) & 1.568 & 1.106 & 2.674 & 0.462 & 1.42 & 0.595 \\
\bottomrule
\end{tabular}
\begin{tablenotes}[flushleft]
\footnotesize
\item \textit{Notes:} As for Table~\ref{tab:main}, without the survey weights. Pooled values are not directly comparable with annual values.
\item \textit{Source:} Authors' calculations based on PLFS person-level data.
\end{tablenotes}
\end{threeparttable}
\end{table}

\begin{table}[htbp]
\centering
\begin{threeparttable}
\caption{Two-sample Kolmogorov--Smirnov tests with conventional p-values}\label{tab:ks}
\small
\begin{tabular}{@{}lcccc@{}}
\toprule
 & \multicolumn{2}{c}{Survey-weighted} & \multicolumn{2}{c}{Unweighted} \\
\cmidrule(lr){2-3}\cmidrule(l){4-5}
Year & $D$ & p-value & $D$ & p-value \\
\midrule
2022 & 0.5648568 & $1.813618\times10^{-157}$ & 0.5716191 & $9.881313\times10^{-324}$ \\
2023 & 0.6375953 & $8.450177\times10^{-199}$ & 0.6409640 & $4.940656\times10^{-324}$ \\
2024 & 0.5770324 & $2.784669\times10^{-163}$ & 0.5908784 & $9.881313\times10^{-324}$ \\
2025 & 0.5354449 & $2.013157\times10^{-143}$ & 0.5920982 & $4.940656\times10^{-324}$ \\
2022--2025 (pooled) & 0.5802397 & 0 & 0.5950764 & $4.940656\times10^{-324}$ \\
\bottomrule
\end{tabular}
\begin{tablenotes}[flushleft]
\footnotesize
\item \textit{Notes:} $H_0$: men and women have the same wage distribution. The p-values are conventional and are not adjusted for the survey design; they are reported for reference only, and $D$ is interpreted descriptively. Unweighted p-values of $4.940656\times10^{-324}$ and $9.881313\times10^{-324}$ are the smallest positive numbers representable in double-precision arithmetic, i.e.\ zero to machine precision. Wages are heaped at round values, so the p-values are also approximate in this respect.
\item \textit{Source:} Authors' calculations based on PLFS person-level data.
\end{tablenotes}
\end{threeparttable}
\end{table}

\begin{table}[htbp]
\centering
\begin{threeparttable}
\caption{Directional Kullback--Leibler divergences by state, survey-weighted}\label{tab:states_w}
\small
\begin{tabular}{@{}lcccccccc@{}}
\toprule
 & \multicolumn{2}{c}{Delhi} & \multicolumn{2}{c}{West Bengal} & \multicolumn{2}{c}{Maharashtra} & \multicolumn{2}{c}{Tamil Nadu} \\
\cmidrule(lr){2-3}\cmidrule(lr){4-5}\cmidrule(lr){6-7}\cmidrule(l){8-9}
Year & $M\|F$ & $F\|M$ & $M\|F$ & $F\|M$ & $M\|F$ & $F\|M$ & $M\|F$ & $F\|M$ \\
\midrule
2022 & 1.508 & 0.304 & 1.592 & 0.820 & 1.354 & 0.954 & 2.789 & 1.568 \\
2023 & 2.813 & 0.689 & 1.553 & 1.008 & 2.155 & 1.166 & 2.481 & 1.970 \\
2024 & 5.063 & 1.576 & 0.968 & 0.465 & 1.174 & 1.267 & 2.996 & 1.829 \\
2025 & 2.778 & 1.160 & 0.528 & 0.422 & 1.475 & 1.077 & 3.022 & 1.718 \\
\addlinespace
Pooled & 1.034 & 0.837 & 0.707 & 0.606 & 1.495 & 1.084 & 2.522 & 1.780 \\
\bottomrule
\end{tabular}
\begin{tablenotes}[flushleft]
\footnotesize
\item \textit{Notes:} Divergences in nats, computed as in Table~\ref{tab:main} within each state; columns $M\|F$ and $F\|M$ report $\DMF$ and $\DFM$.
\item \textit{Source:} Authors' calculations based on PLFS person-level data.
\end{tablenotes}
\end{threeparttable}
\end{table}

\begin{table}[htbp]
\centering
\begin{threeparttable}
\caption{Directional Kullback--Leibler divergences by state, unweighted}\label{tab:states_u}
\small
\begin{tabular}{@{}lcccccccc@{}}
\toprule
 & \multicolumn{2}{c}{Delhi} & \multicolumn{2}{c}{West Bengal} & \multicolumn{2}{c}{Maharashtra} & \multicolumn{2}{c}{Tamil Nadu} \\
\cmidrule(lr){2-3}\cmidrule(lr){4-5}\cmidrule(lr){6-7}\cmidrule(l){8-9}
Year & $M\|F$ & $F\|M$ & $M\|F$ & $F\|M$ & $M\|F$ & $F\|M$ & $M\|F$ & $F\|M$ \\
\midrule
2022 & 1.331 & 8.228 & 2.260 & 0.986 & 1.622 & 1.075 & 4.056 & 1.698 \\
2023 & n.a. & n.a. & 1.521 & 0.863 & 2.076 & 1.186 & 3.663 & 2.289 \\
2024 & 1.228 & 7.663 & 1.257 & 0.606 & 1.852 & 1.253 & 3.997 & 1.832 \\
2025 & 0.837 & 2.428 & 1.208 & 0.872 & 1.650 & 1.180 & 4.015 & 1.765 \\
\addlinespace
Pooled & 0.701 & 1.281 & 1.042 & 0.804 & 1.580 & 1.165 & 3.101 & 1.835 \\
\bottomrule
\end{tabular}
\begin{tablenotes}[flushleft]
\footnotesize
\item \textit{Notes:} As for Table~\ref{tab:states_w}, without the survey weights. For Delhi, the estimate for 2023 is not available (R returned NA), and the reversal of the asymmetry, with large values of $\DFM$ in 2022 and 2024, is likely to reflect the very small number of women in the unweighted Delhi sample; the Delhi unweighted estimates are not used for inference. State-level estimates in general require a minimum effective sample-size criterion before they can be interpreted.
\item \textit{Source:} Authors' calculations based on PLFS person-level data.
\end{tablenotes}
\end{threeparttable}
\end{table}

\clearpage
\section{Unweighted wage densities}\label{app:udens}
\setcounter{figure}{0}

\begin{figure}[htbp]
\centering
\includegraphics[width=0.72\textwidth]{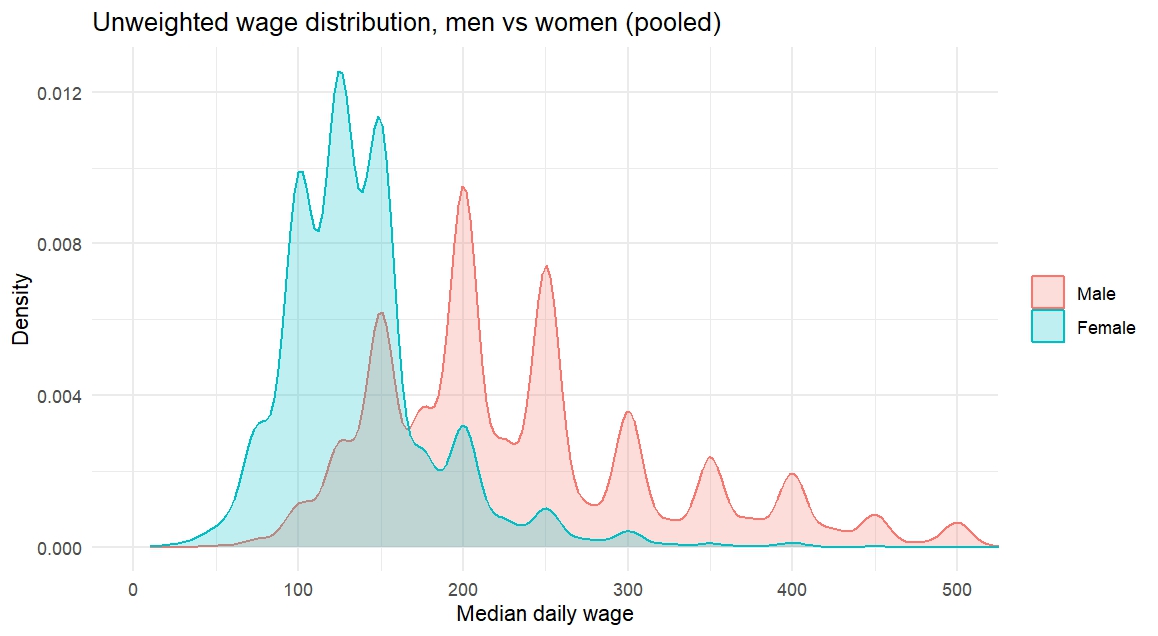}
\caption{Unweighted densities of median daily casual wages, men and women, 2022--2025 pooled}\label{fig:udens_pooled}
\begin{minipage}{0.85\textwidth}\footnotesize
\textit{Notes:} Gaussian kernel density estimates without survey weights; wages in rupees per day.\\
\textit{Source:} Authors' calculations based on PLFS person-level data.
\end{minipage}
\end{figure}

\begin{figure}[htbp]
\centering
\includegraphics[width=0.76\textwidth]{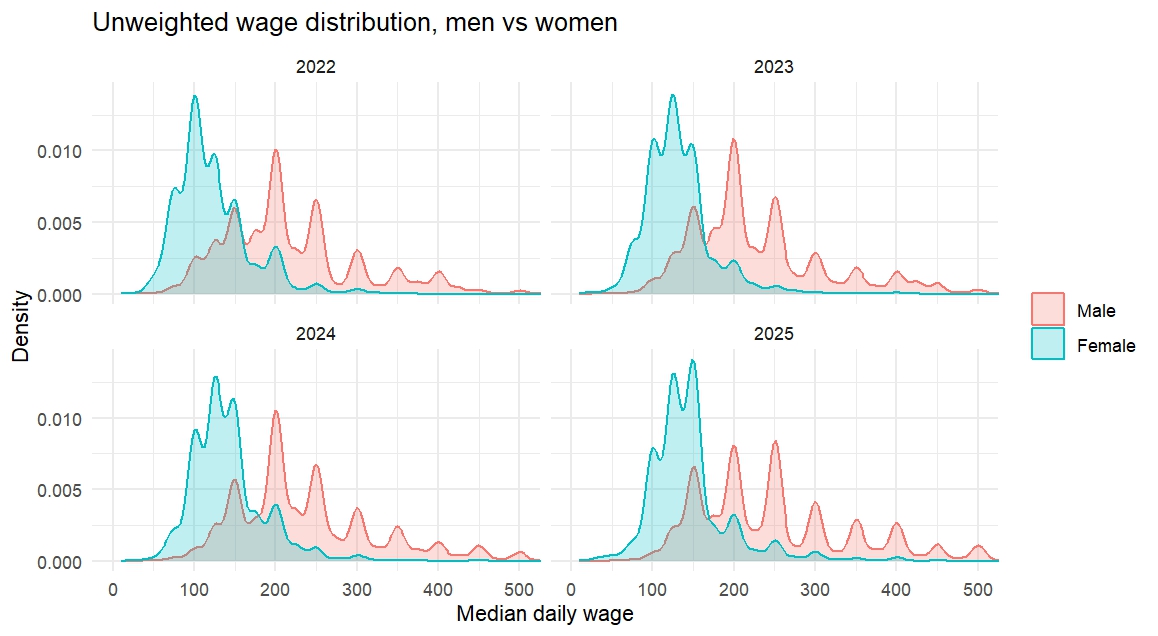}
\caption{Unweighted densities of median daily casual wages, men and women, by year}\label{fig:udens_year}
\begin{minipage}{0.9\textwidth}\footnotesize
\textit{Notes:} As for Figure~\ref{fig:udens_pooled}.\\
\textit{Source:} Authors' calculations based on PLFS person-level data.
\end{minipage}
\end{figure}

\clearpage
\section{Quantile--quantile plots}\label{app:figs}
\setcounter{figure}{0}

\begin{figure}[htbp]
\centering
\includegraphics[width=0.62\textwidth]{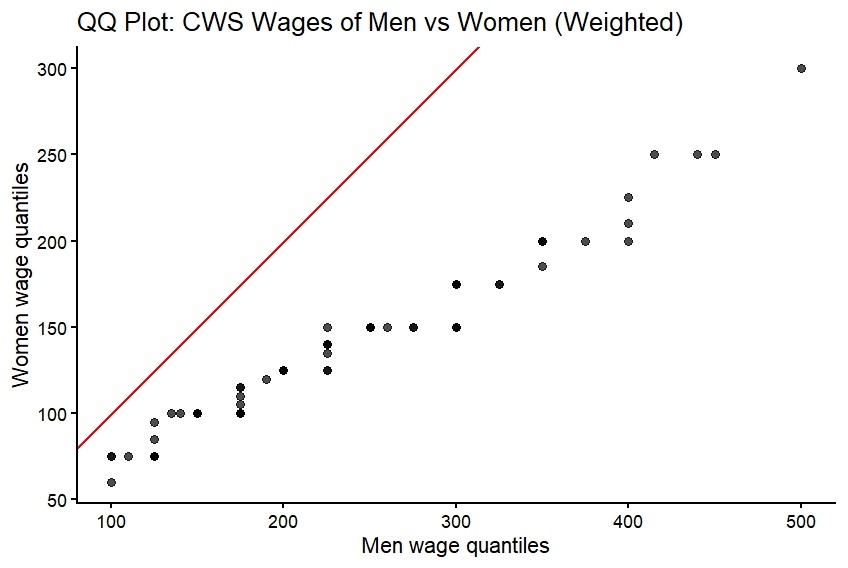}
\caption{Quantile--quantile plot of daily casual wages, women against men, survey-weighted, 2022--2025 pooled}\label{fig:wpooled}
\begin{minipage}{0.66\textwidth}\footnotesize
\textit{Notes:} Each point plots a quantile of the female wage distribution against the same quantile of the male distribution, in rupees per day. The red line is the 45-degree line of equality.\\
\textit{Source:} Authors' calculations based on PLFS person-level data.
\end{minipage}
\end{figure}

\begin{figure}[htbp]
\centering
\includegraphics[width=0.66\textwidth]{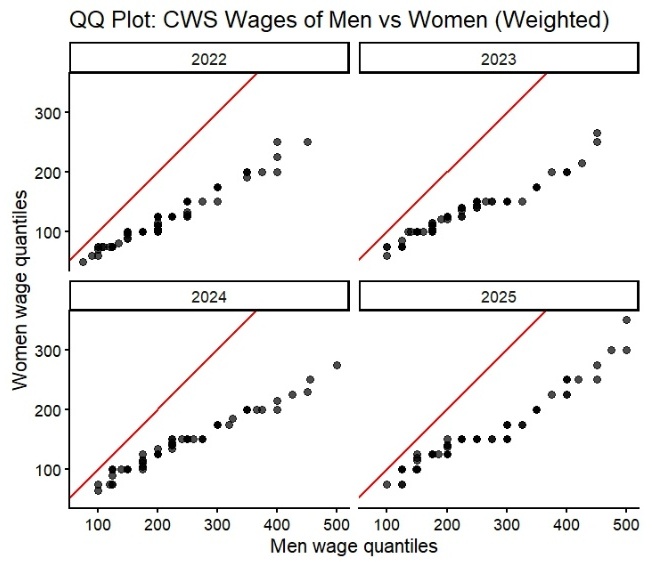}
\caption{Quantile--quantile plots of daily casual wages, women against men, survey-weighted, by year}\label{fig:wyear}
\begin{minipage}{0.66\textwidth}\footnotesize
\textit{Notes:} As for Figure~\ref{fig:wpooled}.\\
\textit{Source:} Authors' calculations based on PLFS person-level data.
\end{minipage}
\end{figure}

\begin{figure}[htbp]
\centering
\includegraphics[width=0.62\textwidth]{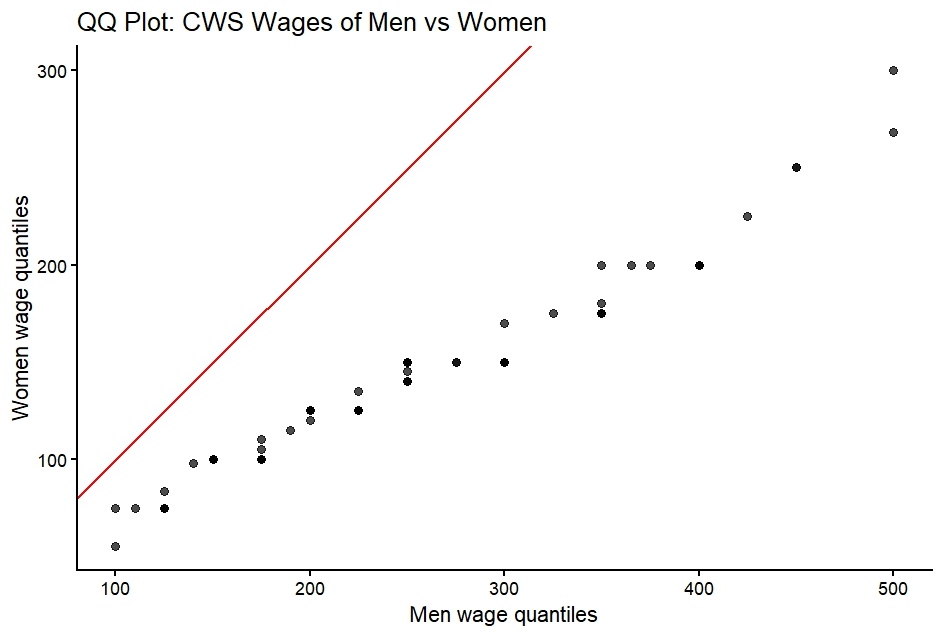}
\caption{Quantile--quantile plot of daily casual wages, women against men, unweighted, 2022--2025 pooled}\label{fig:upooled}
\begin{minipage}{0.66\textwidth}\footnotesize
\textit{Notes:} As for Figure~\ref{fig:wpooled}, without the survey weights.\\
\textit{Source:} Authors' calculations based on PLFS person-level data.
\end{minipage}
\end{figure}

\begin{figure}[htbp]
\centering
\includegraphics[width=0.66\textwidth]{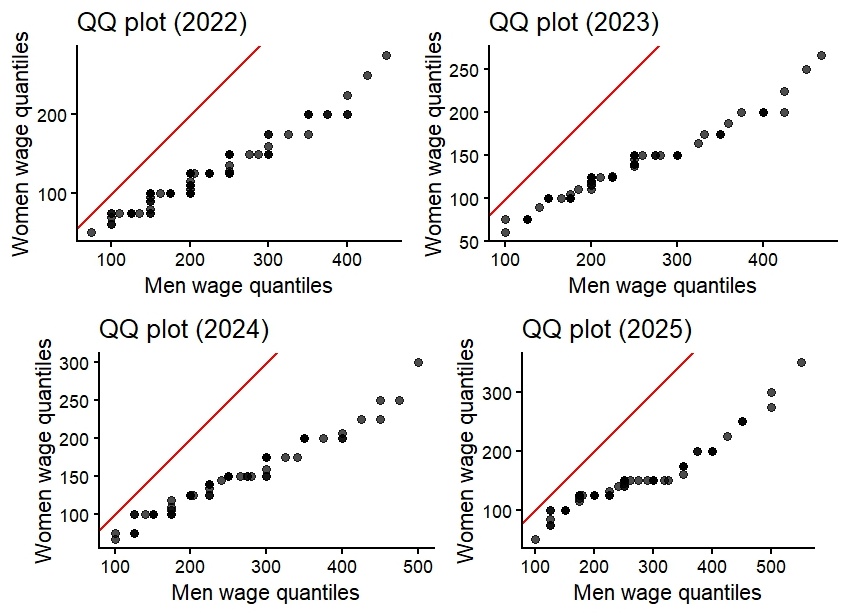}
\caption{Quantile--quantile plots of daily casual wages, women against men, unweighted, by year}\label{fig:uyear}
\begin{minipage}{0.66\textwidth}\footnotesize
\textit{Notes:} As for Figure~\ref{fig:wpooled}, without the survey weights.\\
\textit{Source:} Authors' calculations based on PLFS person-level data.
\end{minipage}
\end{figure}

Every point lies below the 45-degree line in every year, so female quantiles are lower than male quantiles throughout. In the weighted plots, female quantiles are roughly two-thirds of male quantiles or more in the lower part of the distribution and about one-half to three-fifths of them where male quantiles lie between \rupee300 and \rupee500, so the proportional gap is larger in the upper half. Horizontal runs at round values such as \rupee100, \rupee125, \rupee150 and \rupee200 reflect heaping of female wages. The unweighted plots (Figures~\ref{fig:upooled} and~\ref{fig:uyear}) show the same pattern.

\end{document}